\documentclass[%
aps,
prl,
twocolumn,
groupedaddress,
]{revtex4-2}

\usepackage{graphicx}
\usepackage{dcolumn}
\usepackage{bm}
\usepackage{chemformula}
\usepackage{siunitx}
\usepackage{romannum}
\usepackage{amsfonts}

\begin{document}

\preprint{APS/123-QED}

\title{\textbf{Degenerate high-order hybrid bound states in the continuum beyond diffraction limit}
}%

\author{Ji Tong Wang}
\email[e-mail: ]{jitong.wang@ucl.ac.uk}
\affiliation{%
  Department of Electronic and Electrical Engineering,\\
  University College London, Torrington Place, London WC1E 7JE, United Kingdom
}

\author{Nicolae C. Panoiu}
\email[e-mail: ]{n.panoiu@ucl.ac.uk}
\affiliation{%
  Department of Electronic and Electrical Engineering,\\
  University College London, Torrington Place, London WC1E 7JE, United Kingdom
}

\date{\today}

\begin{abstract}
We demonstrate the existence of at-$\Gamma$ degenerate hybrid bound-states in the continuum (BICs) with high-order topological charges and quadratic band degeneracy in periodic photonic structures with $C_{6v}$ symmetry above the diffraction limit. The degenerate BICs are realized by combining symmetry protection, which suppresses radiation into zeroth-order diffraction channel, with quadratic band degeneracy and parameter tuning to eliminate radiation into first-order diffraction channels. An effective Hamiltonian is used to investigate the band structure around the degeneracy point, and the topological dynamics in all diffraction channels are characterized. This work reveals a novel mechanism for realizing degenerate BICs above the diffraction limit, opening up opportunities for new physics and applications.
\end{abstract}

\maketitle


The concept of topological singularities has been widely studied in wave systems, which are characterized by nontrivial winding patterns of the phase or polarization of fields in real space \cite{Freund2002}. In the context of photonic crystals (PhCs), recent studies suggest that far-field polarization vector fields radiated from such structures exhibit intriguing topological structures in momentum space \cite{Zhen2014}. These momentum-space polarization singularities include vortices (\textit{V} points) characterized by vanishing far-field intensity and integer topological charges, circularly polarized states (\textit{C} points) with half-integer topological charges, and band degenerate points with undetermined far-field polarization. Photonic band degeneracies are commonly classified by their local band dispersion as linear Dirac points, characterized by half-integer topological charge, and quadratic degeneracy (\textit{QD}) points carrying integer topological charge \cite{Chen2019}. In particular, optical bound states in the continuum (BICs), which manifest as \textit{V} points in the far-field polarization distribution, have attracted significant attention due to their non-radiating nature despite being embedded in the continuum spectrum \cite{Kang2023}. This enables key applications in sensing \cite{Tittl2018}, lasing \cite{Ratiu2025}, and nonlinear optics \cite{Minkov2019,Wang2025}. Moreover, the conversion among different types of singularities, constrained by symmetry and total topological charge conservation, provides a powerful mechanism for exploring novel physics, such as BICs evolving from or into pairs of \textit{C} points \cite{Liu2019,Kang2025} and high-order BICs formed through the coalescence of accidental BICs and \textit{QD} points \cite{Su2026}.

Most previous studies of momentum-space polarization singularities in PhCs focused on BICs located below the diffraction limit \cite{Kang2023}. Recently, nondegenerate at-$\Gamma$ hybrid BICs (h-BICs) were introduced as a generic concept to study \textit{V} points and \textit{C} points above this frequency cutoff \cite{Wang2026}, whereby symmetry protection and parameter tuning were used to suppress radiation. However, $\Gamma$-point degenerate BICs supporting multiple diffraction channels remain unexplored. In particular, band degeneracy can naturally give rise to nontrivial polarization winding in momentum space \cite{Chen2019}, leading to qualitatively distinct topological singularities and radiation characteristics that are absent in nondegenerate modes. These features point to at-$\Gamma$ degenerate modes as potentially representing an effective platform for exploring radiation and polarization topology across multiple diffraction channels above the diffraction limit.

In this Letter, we propose a scheme to realize at-$\Gamma$ degenerate h-BICs with high-order topological charges in all diffraction channels of a PhC slab with hexagonal lattice ($C_{6v}$ symmetry). To realize such h-BICs beyond the diffraction limit, we utilize symmetry protection to suppress radiation into zeroth-order channel, and combine quadratic band degeneracy with structural parameter tuning of the PhC to eliminate light emission into the first-order channels. We demonstrate the existence of degenerate h-BICs in the geometric parameter space and reveal the formation of high-order \textit{V} points in first-order channels by merging \textit{QD} points with pairs of \textit{C} points \textit{via} tuning of the geometrical parameters. The evolution of polarization singularities is governed by global topological charge conservation. Moreover, an effective Hamiltonian model is employed to characterize the band structure and scaling of the quality ($Q$)-factor in the vicinity of the degenerate point.

\begin{figure}[t]
\centering
\includegraphics[width=\columnwidth]{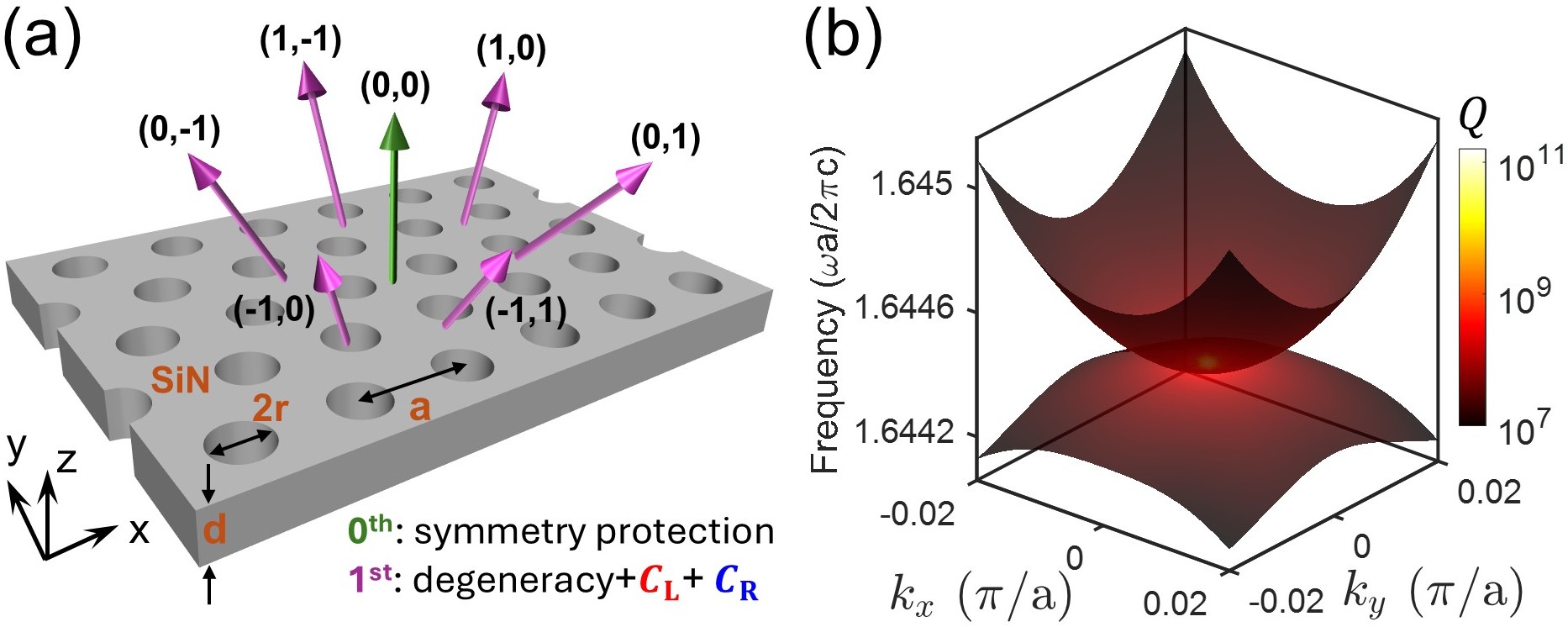}
\caption{(a) Schematic of a PhC slab with hexagonal lattice supporting zeroth- and first-diffraction orders and formation mechanisms of at-$\Gamma$ degenerate h-BICs. $C_L$ ($C_R$) denotes left- (right-) handed $C$ points carrying the same topological charge. (b) Band surfaces and corresponding $Q$-factor distribution of the TM-like, at-$\Gamma$ degenerate modes belonging to the $E_2$ representation and corresponding to $(\tilde{r},\tilde{d})=(0.226,0.4715)$. The color map represents the $Q$-factor of the modes.}
\label{fig1}
\end{figure}

The schematics of the PhC slab with hexagonal lattice investigated in this work are presented in Fig.~\ref{fig1}(a). The free-standing PhC slab, assumed to be made of SiN with refractive index $n=1.98$, consists of an array of air holes with radius $r$. The lattice constant is $a=\SI{995}{\nm}$ and the slab thickness is $d$. Due to the up-down mirror symmetry ($\sigma_z$), all modes of the PhC slab can be separated into transverse electric (TE)-like and transverse magnetic (TM)-like modes. Here, we focus on the TM-like modes. At the $\Gamma$-point, the normalized frequencies of the first- and second-order diffraction limits are $\tilde{\omega}_{I}=2/\sqrt{3}$ and $\tilde{\omega}_{II}=2$, respectively, where $\tilde{\omega}=\omega a/(2\pi c)$ is the normalized frequency, with $\omega$ and $c$ being the optical frequency and the speed of light in vacuum, respectively. For at-$\Gamma$ degenerate modes (\textit{QD} points) that lie between $\tilde{\omega}_{I}$ and $\tilde{\omega}_{II}$, for both modes there exist seven diffraction channels (one zeroth-order and six first-order channels), as shown in Fig.~\ref{fig1}(a). For each channel defined by a pair of integers $(m_1,m_2)$, the far-field of degenerate modes possess the same wavevector, $\mathbf{k}_{m_1m_2} = (\mathbf{k}_{\parallel}+\mathbf{G}_{m_1m_2}, k_z = \sqrt{n_0^2\omega^2/c^2 - \vert\mathbf{k}_{\parallel}+\mathbf{G}_{m_1m_2}\vert^2})$, where $n_0$ is the refractive index of the environment, $\mathbf{k}_{\parallel}$ is the in-plane wavevector in the first Brillouin zone, and $\mathbf{G}_{m_1m_2}$ is the corresponding reciprocal lattice vector. Due to the $C_{6v}$ symmetry, the degenerate modes have the same frequencies and total radiative leakage rates. However, while the leakage rates into the symmetry-invariant zeroth-order channels are identical, the leakage rate into a specific first-order channel can differ between the two modes.

At the $\Gamma$-point, radiation in the zeroth-order channels transforms according to the $E_1$ irreducible representation of the $C_{6v}$ point group \cite{Sakoda2005}. Therefore, we focus on degenerate modes with representation $E_2$, for which symmetry mismatch forbids coupling to zeroth-order channels and gives rise to $V$ points. Specifically, at the $\Gamma$-point degenerate $V$ points, the vanishing zeroth-order far-field is protected by the $C_2$ symmetry of the $E_2$ modes, whereas the quadratic band degeneracy is due to the $C_3$ symmetry \cite{Doiron2024}. By contrast, parameter tuning is required to suppress radiation into first-order diffraction channels because symmetry alone generally cannot suppress all open diffraction channels above the diffraction limit \cite{Cerjan2021}. Thus, at \textit{QD} points, first-order channels are generally radiative. However, the far-field polarization is not uniquely defined due to degeneracy, giving rise to integer charges at the QD points \cite{Zhang2018}. By tuning the geometric parameters $r$ and $d$, the simultaneous formation of a $V$ point with a high-order charge can be realized when pairs of $C$ points coalesce with the degeneracy point in each first-order channel of the degenerate modes. Thus, as illustrated in Fig.~\ref{fig1}(b), by tuning the geometric parameters, the PhC slab with $(\tilde{r},\tilde{d}) = (0.226,0.4715)$, where $\tilde{r}\equiv r/a$ and $\tilde{d}\equiv d/a$, is found to support a pair of at-$\Gamma$ TM-like doubly degenerate h-BICs with representation $E_2$. Using the finite-element method \cite{COMSOL,RSOFT}, the computed normalized frequency is $\tilde{\omega}_{BIC}=1.6445$ $(\tilde{\omega}_{II}>\tilde{\omega}_{BIC}>\tilde{\omega}_{I})$, and the $Q$-factors diverge at the QD point.

\begin{figure}[t]
\centering
\includegraphics[width=\columnwidth]{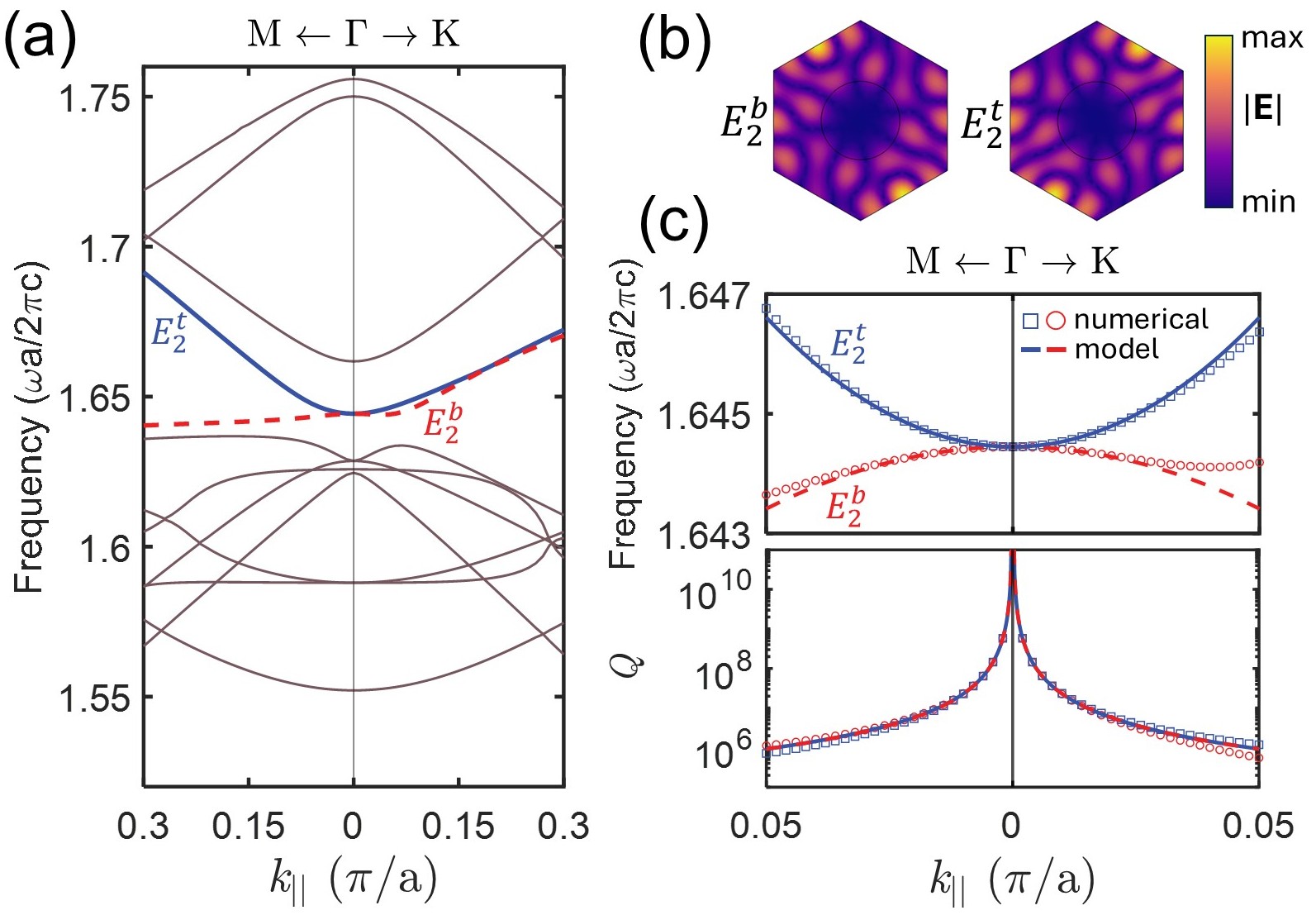}
\caption{(a) Band structure for the TM-like modes of a PhC slab with hexagonal lattice with $(\tilde{r},\tilde{d}) = (0.226,0.4715)$. The bottom (top) band of the modes degenerate at $\Gamma$, denoted as $E_2^b$ ($E_2^t$), is shown as red dashed (blue solid) lines. (b) Electric field profiles in the middle plane of the PhC slab at the $\Gamma$ point. (c) Band dispersion (top panel) and $Q$-factor (bottom panel) of at-$\Gamma$ degenerate modes nearby the degeneracy point. The results obtained from the effective Hamiltonian model are shown as red and blue lines, whereas those from numerical simulations are marked by red circles and blue squares.}
\label{fig2}
\end{figure}

In Fig.~\ref{fig2}(a), we show the photonic band structure for the two TM-like modes, which are degenerate at the $\Gamma$ point, in the vicinity of the $\Gamma$ point. The calculations are performed for the PhC slab with $(\tilde{r},\tilde{d}) = (0.226,0.4715)$, for which the two modes are BICs at the $\Gamma$ point. The at-$\Gamma$ electric field distributions in the middle plane of the PhC slab are plotted in Fig.~\ref{fig2}(b) for the two BICs. Away from the $\Gamma$-point, the degeneracy is lifted and the two bands become quasi-BICs with finite $Q$-factors, as per Fig.~\ref{fig1}(b), whereby the bottom ($E_2^b$) and top ($E_2^t$) bands are shown. Along the high-symmetry directions away from $\Gamma$, the $s$- and $p$-polarized incident waves couple only to a single band according to the symmetry selection rules \cite{Sakoda2005}, highlighting the polarization-selective excitation of the two modes.

To gain deeper insights into the dispersion and radiation behaviors of the pair of $E_2$ modes near the $\Gamma$-point, we utilize the symmetry-constrained effective Hamiltonian for PhC slabs with $C_{6v}$ symmetry derived from the $\bm{k} \cdot \bm{p}$ theory as \cite{Chong2008,Long2023}:
\begin{equation}
\begin{aligned}
\tilde{H}_{\mathrm{eff}}&=(\tilde{\omega}_0-i\tilde{\gamma}_0)\mathbf{I}\\
&\quad+\begin{pmatrix}
\alpha|\tilde{\mathbf{k}}_{\parallel}|^2+\beta(\tilde{k}_x^2-\tilde{k}_y^2) & 2\beta \tilde{k}_x \tilde{k}_y\\
2\beta \tilde{k}_x \tilde{k}_y & \alpha|\tilde{\mathbf{k}}_{\parallel}|^2-\beta(\tilde{k}_x^2-\tilde{k}_y^2)
\end{pmatrix},
\end{aligned}
 \label{equ1}
\end{equation}
where $\alpha$ and $\beta$ are complex coefficients, $\tilde{\omega}_0$ and $\tilde{\gamma}_0=\gamma_0a/(2\pi c)$ are the normalized resonant frequency and decay rate at the degeneracy point, respectively, and $\tilde{k}_{x,y}=k_{x,y}a/\pi$ are normalized wavevnumber components. The real (imaginary) part of the eigenvalues of the effective Hamiltonian governs the frequency (leakage rate) dispersion. This Hamiltonian has two eigenvalues
\begin{equation}
\tilde{\omega}_{\pm}(\tilde{\mathbf{k}}_{\parallel}) - i\tilde{\gamma}_{\pm}(\tilde{\mathbf{k}}_{\parallel})
= \tilde{\omega}_0 - i\tilde{\gamma}_0 + (\alpha \pm \beta)\,|\tilde{\mathbf{k}}_{\parallel}|^2.
\label{equ2}
\end{equation}
Here, the $``+''$ and $``-''$ subscripts denote the $E_2^t$ and $E_2^b$ bands, respectively. From this equation, it is evident that the frequency splitting scales quadratically and isotropically near the degeneracy point. Using the band structures along the $\Gamma-\mathrm{M}$ and $\Gamma-\mathrm{K}$ axes, the complex coefficients are fitted as $\alpha = 0.2254 - 3.6\times10^{-4}i$ and $\beta = 0.64 + 6\times10^{-7}i$. Then, the $Q$-factor is calculated as $Q=\tilde{\omega}/(2\tilde{\gamma})$. In Fig.~\ref{fig2}(c) we compare the results for the two bands determined by the numerical method and the effective Hamiltonian, excellent agreement being observed for $\vert\tilde{\mathbf{k}}_{\parallel}\vert < 0.02$. Note also that the two bands exhibit nearly identical $Q$ for $\vert\tilde{\mathbf{k}}_{\parallel}\vert < 0.05$.
\begin{figure}[t]
\centering
\includegraphics[width=\columnwidth]{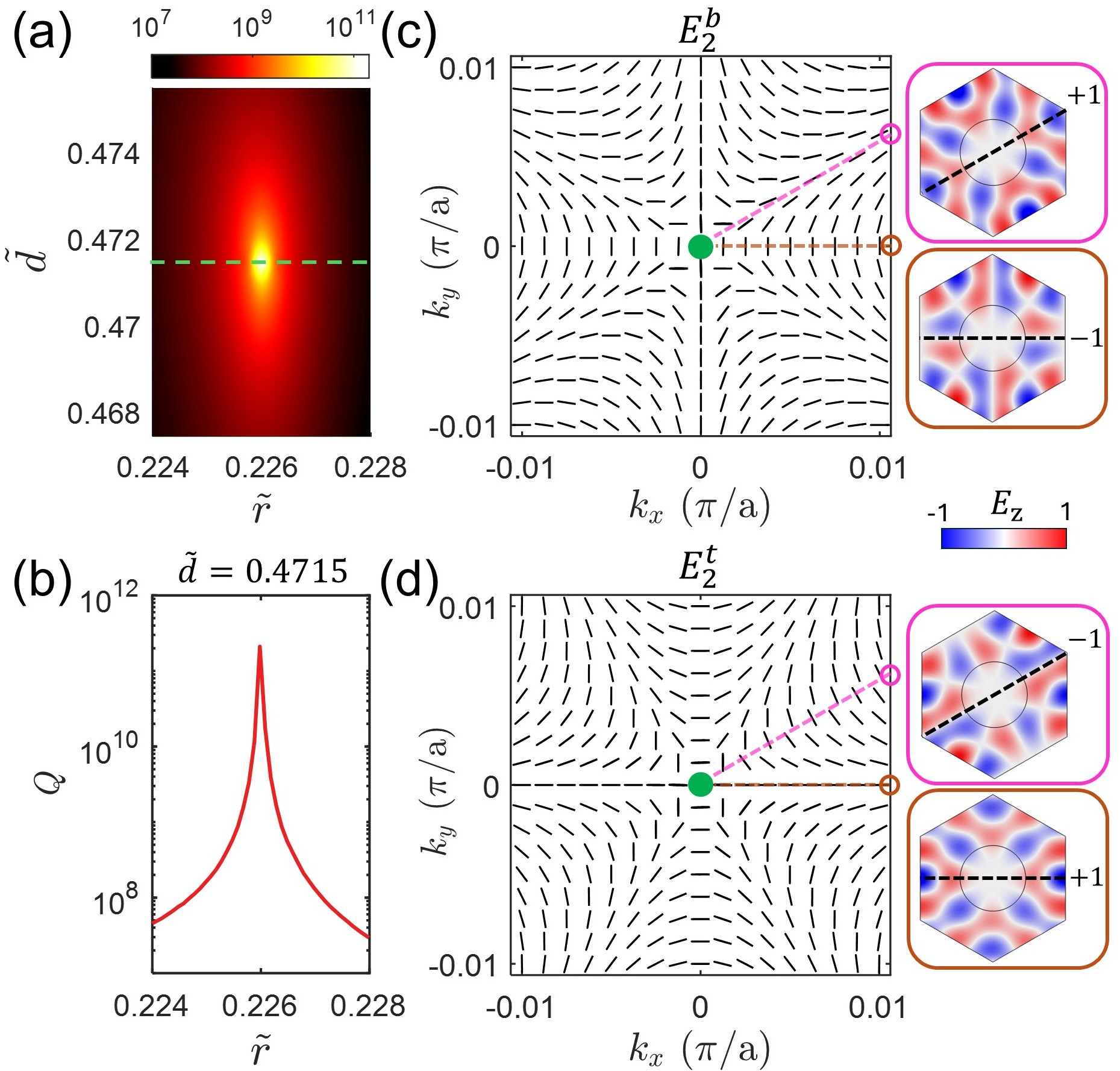}
\caption{(a) Dependence of $Q$ on $\tilde{r}$ and $\tilde{d}$, determined for the at-$\Gamma$ degenerate modes. (b) $Q$ \textit{vs}. $\tilde{r}$, computed for $\tilde{d}=0.4715$, namely along the green dashed line in (a). (c) Left panel: far-field polarization map of the $E_2^b$ mode for (0,0) diffraction channel, where the pink (brown) dashed line denotes the $\Gamma-\mathrm{M}$ ($\Gamma-\mathrm{K}$) axis. The green dot denotes a $V$ point with topological charge of $-2$. Right panel: electric field profiles evaluated at the locations marked by circles in the left panel, with $\pm1$ being the eigenvalues of the indicated mirror symmetry transformations. (d) The same as in (c), but for the $E_2^t$ mode.}
\label{fig3}
\end{figure}

Next, to clarify how geometric parameter tuning governs the emergence of h-BICs, we evaluate the $Q$-factor of the degenerate modes at the $\Gamma$-point as a function of parameters $\tilde{r}$ and $\tilde{d}$, as per Fig.~\ref{fig3}(a). Due to the $C_{6v}$ symmetry that prohibits radiation into zeroth-order channels, the $Q$-factor is determined solely by the total radiated power into all first-order radiative channels. At $(\tilde{r},\tilde{d}) = (0.226,0.4715)$, the $Q$-factor diverges, indicating the existence of a pair of degenerate BICs. To further illustrate the formation of the BICs, we vary $\tilde{r}$ at fixed $\tilde{d}=0.4715$, as indicated in Fig.~\ref{fig3}(a) by the green dashed line. As shown in Fig.~\ref{fig3}(b), as expected, the $Q$-factor diverges at $\tilde{r}=0.226$.

As BICs correspond to polarization singularities in momentum space, their topological properties deserve special attention in the case of degenerate and multi-channel configuration. The topological charge, $q$, can be determined from the winding number of the major axis of the polarization ellipse around the singularity in the momentum space as \cite{Yoda2020}:
\begin{equation}
    q=\frac{1}{2\pi}\oint_{\mathcal{C}} d \mathbf{k}_{\parallel} \cdot\nabla_{\mathbf{k}_{\parallel}}\phi(\mathbf{k}_{\parallel}),
    \label{equ3}
\end{equation}
where $\phi(\mathbf{k}_{\parallel})=\frac{1}{2} \mathrm{arg}[S_1(\mathbf{k}_{\parallel})+iS_2(\mathbf{k}_{\parallel})]$ is the polarization angle, $S_i(\mathbf{k}_{\parallel})$ are the Stokes parameters of the far-field polarization, $\mathbf{d}(\mathbf{k}_{\parallel})$, which is found by projecting the polarization vector in the $s$-$p$ plane onto the $x$-$y$ plane with the polarization ellipse preserved, and $\mathcal{C}$ is a closed path around the singularity.

We first analyze the topology of the zeroth-order channels. The far-field polarization maps computed for (0,0) channels of the $E_2^b$ and $E_2^t$ modes are presented in Figs.~\ref{fig3}(c) and \ref{fig3}(d), respectively, for a PhC slab with $(\tilde{r},\tilde{d}) = (0.226,0.4715)$. As can be seen, the far-field polarization vector of the $E_2^b$ ($E_2^t$) mode is perpendicular (parallel) to the mirror-invariant line of $\Gamma-\mathrm{K}$, but parallel (perpendicular) to the mirror-invariant line of $\Gamma-\mathrm{M}$. Due to the sixfold symmetry, the allowed topological charge of the degenerate $V$ points at the $\Gamma$-point is $q_{00} = 6n-2$ \cite{Zhang2018}, with $n=0$ here. In addition, Figs.~\ref{fig3}(c) and \ref{fig3}(d) show the $E_z$ profiles along the high-symmetry directions, with $\pm1$ denoting the eigenvalues of the mirror symmetry transformation.
\begin{figure}[t]
\centering
\includegraphics[width=\columnwidth]{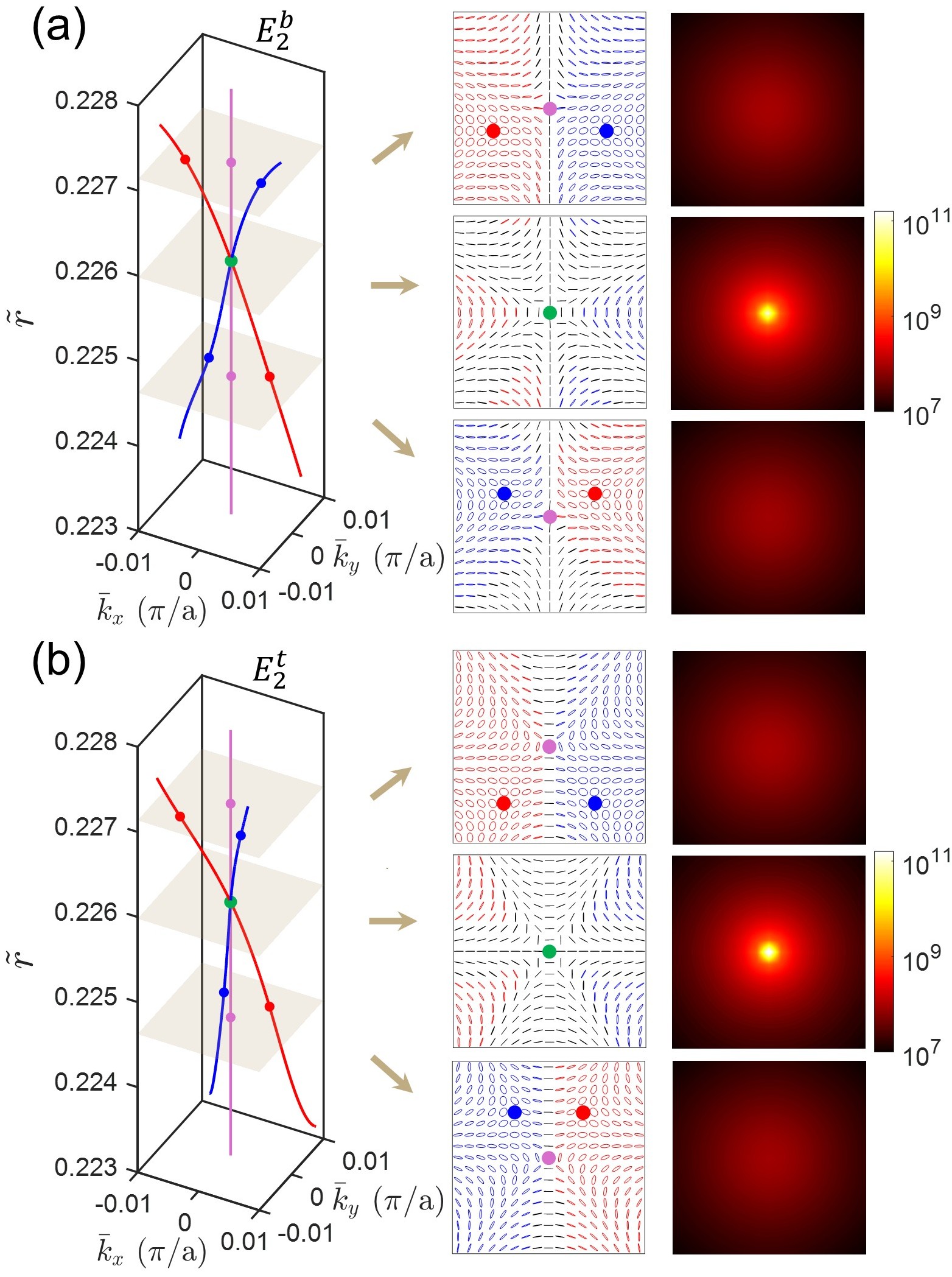}
\caption{$\mathbf{k}_{\parallel}$-space trajectories of $C$ and $QD$ points in $(1,-1)$ channel for (a) $E_2^b$ mode and (b) $E_2^t$ mode when $\tilde{r}$ is varied at fixed $\tilde{d}=0.4715$. Red (blue) dots denote $C$ points with $-1/2$ charge and left (right) handedness, whereas purple dots denote $QD$ points with charge of $-1$. Middle and right columns show the maps of far-field polarization vector in $(1,-1)$ channel and total $Q$-factor, respectively, determined at $\tilde{r}$ = \numlist[list-final-separator={, and }]{0.2246;0.226;0.2271} in the range $\bar{k}_{x}, \bar{k}_{y}\in[-0.01,0.01]\,\pi/a$.}
\label{fig4}
\end{figure}

Regarding the first-order diffraction channels, Fig.~\ref{fig3}(a) highlights the role of parameter tuning in the formation of degenerate BICs, which can be further understood from a topological perspective. Due to $C_{6v}$ symmetry, it is enough to consider for both $E_2^b$ and $E_2^t$ only one first-order diffraction channel, say $(1,-1)$, to study the polarization singularities dynamics. To this end, the shifted wavevector $(\bar{k}_x,\bar{k}_y)=(k_x,k_y-4\pi/\sqrt{3}a)$ is introduced.

The left panels of Fig.~\ref{fig4} show the $\mathbf{k}_{\parallel}$-space trajectories of polarization singularities as the hole radius $\tilde{r}$ varies from \numrange{0.223}{0.228} at fixed $\tilde{d}=0.4715$. For both degenerate modes, a singular $QD$ point with charge of $-1$ is pinned at $(\bar{k}_x,\bar{k}_y)=(0,0)$ due to preserved symmetry. In addition, a pair of $C$ points related by mirror symmetry $\sigma_x$ with charge of $-1/2$ and opposite handedness move towards $(\bar{k}_x,\bar{k}_y)=(0,0)$ and merge with the $QD$ point when $\tilde{r} = 0.226$, thereby forming a $V$ point with charge of $-2$ and vanishing far-field intensity. The corresponding far-field polarization distributions clearly show the winding pattern around the polarization singularity. It should be emphasized that this merging process occurs simultaneously in the $(1,-1)$ channels of both $E_2^b$ and $E_2^t$ modes, and the $C_{6v}$ symmetry of the PhC ensures the simultaneous formation of $V$ points with charge of $-2$ in all first-order channels. Therefore, degenerate h-BICs beyond the diffraction limit form for $(\tilde{r},\tilde{d}) = (0.226,0.4715)$, which can also be inferred from the $Q$-maps presented in the right panels of Fig.~\ref{fig4}. Further increasing $\tilde{r}$ causes the $V$ point to split into two $C$ points with charge of $-1/2$ and a $QD$ point with charge of $-1$. Moreover, for all first-order channels, the locations of the $C$ points in the $\mathbf{k}_{\parallel}$-space are different for the two modes because of the lifted degeneracy away from $\Gamma$. Throughout this evolution of the singularities, the total topological charge is conserved.
\begin{figure}[t]
\centering
\includegraphics[width=\columnwidth]{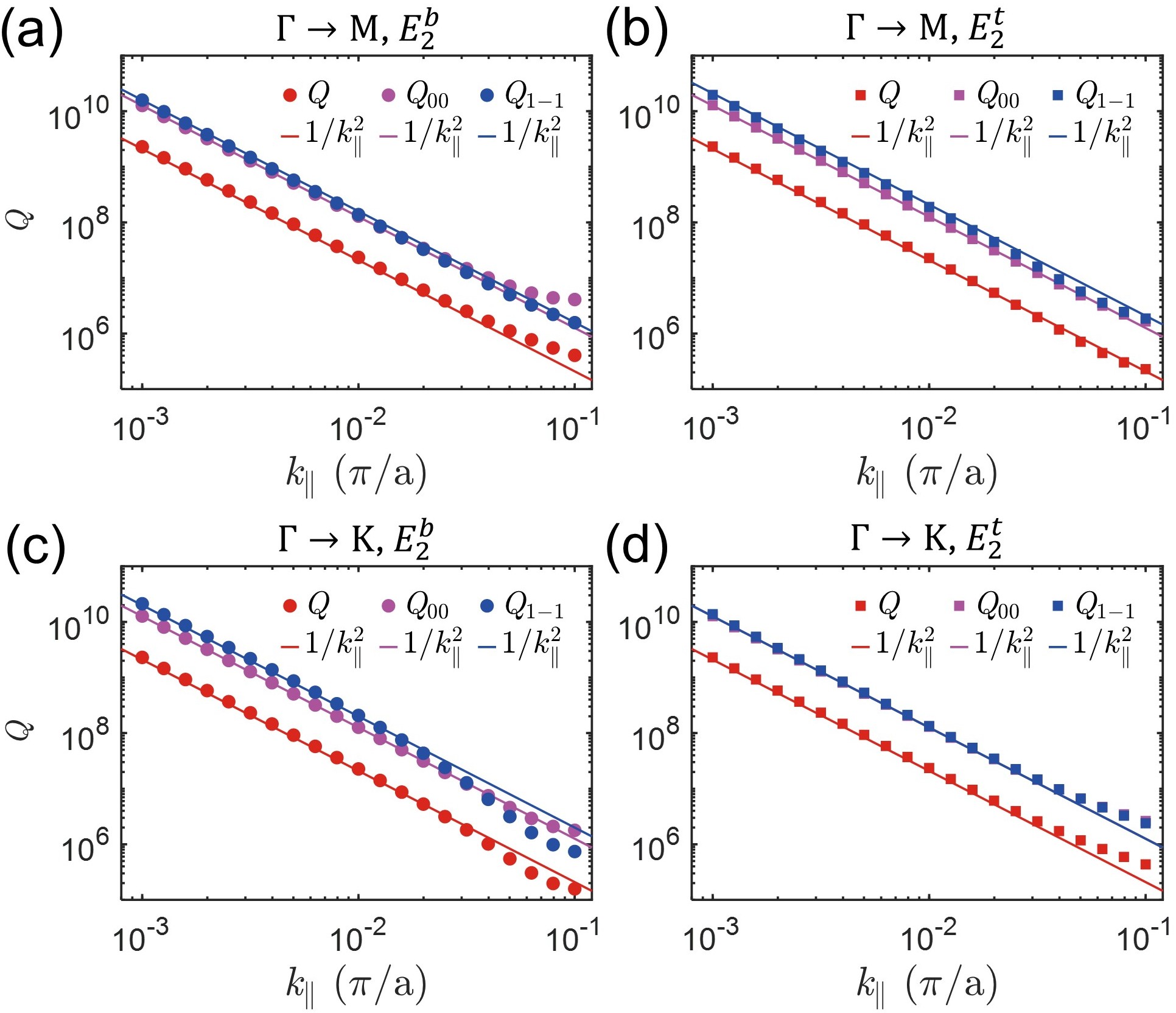}
\caption{Scaling of $Q$-factor along the $\Gamma-\mathrm{M}$ direction for (a) $E_2^b$ mode and (b) $E_2^t$ mode. Scaling of $Q$-factor along the $\Gamma-\mathrm{K}$ direction for (c) $E_2^b$ mode and (d) $E_2^t$ mode. The dots and squares represent the numerically calculated values at sampled $k_{\parallel}$-points, whereas the curves represent the fitting of these results with the $1/k_{\parallel}^2$ function.}
\label{fig5}
\end{figure}

To further characterize the underlying physics of the $E_2$-type degenerate BICs with $\gamma_0=0$, we use the effective Hamiltonian model given in \eqref{equ1} to determine the scaling of the $Q$-factor in nearby the degeneracy point. To this end, we find
\begin{equation}
    Q_{\pm} = -\frac{\tilde{\omega}_0 + \mathfrak{Re}(\alpha \pm \beta)|\tilde{\mathbf{k}}_{\parallel}|^2}{2\mathfrak{Im}(\alpha \pm \beta)|\tilde{\mathbf{k}}_{\parallel}|^2} \approx -\frac{\tilde{\omega}_0}{2\mathfrak{Im}(\alpha \pm \beta)}\cdot \frac{1}{|\tilde{\mathbf{k}}_{\parallel}|^2},
\label{equ4}
\end{equation}
where $\mathfrak{Re}(z)/\mathfrak{Im}(z)$ is the real/imaginary part of the complex number $z$. This equation indicates that the total $Q$-factors scale isotropically as $1/\vert\tilde{\mathbf{k}}_{\parallel}\vert^2$ around the degeneracy point, with the scaling coefficients $\tilde{\omega}_0/[-2\mathfrak{Im}(\alpha \pm \beta)]$. In Fig.~\ref{fig5}, we present the numerically calculated total $Q$-factors at different $\mathbf{k}_{\parallel}$-points along the $\Gamma-\mathrm{M}$ and $\Gamma-\mathrm{K}$ axes, with a good agreement with the fitted results for $\vert\mathbf{k}_{\parallel}\vert<0.1\,\pi/a$ being observed. Noted that, although the degenerate BICs exhibit $V$ points with charge of $-2$ in all channels, the scaling of the total $Q$-factor differs from that of previously reported nondegenerate BICs below the diffraction limit, whereby a topological charge of $q=\pm n$ leads to $Q \propto 1/\vert\mathbf{k}_{\parallel}\vert^{2\vert n\vert}$ \cite{Jin2019}. Moreover, for all zeroth- and first-order channels, the corresponding channel-resolved $Q_{m_1m_2}$ are found to follow the scaling law of $1/|\mathbf{k}_{\parallel}|^2$, with scaling coefficients varying slightly among channels. To illustrate this fact, $Q_{00}$ and $Q_{1-1}$ are shown in Fig.~\ref{fig5}. This also validates the scaling law of the total $Q$-factor since $1/Q = \sum_{m_1,m_2} 1/Q_{m_1,m_2}$.

In summary, we propose a novel scheme to construct at-$\Gamma$ degenerate h-BICs with high-order charges beyond the diffraction limit of periodic photonic structures with $C_{6v}$ symmetry. Symmetry protection suppresses radiation for zeroth-diffraction orders, whereas parameter tuning is used to suppress radiation for first-diffraction orders through the merging of $QD$ points with pairs of $C$ points. The $C_{6v}$ symmetry robustly ensures the simultaneous generation of $V$ points in all diffraction channels of the degenerate h-BICs. More generally, our findings demonstrate that band degeneracy provides a mechanism for realizing multi-channel h-BICs and nontrivial polarization topology above the diffraction limit. This work opens up opportunities for topology tailoring, high-$Q$ photonics, polarization control, and multiplexed diffraction engineering.

\noindent\textit{Acknowledgments}---This work was supported in part by the UK's Engineering and Physical Sciences Research Council under Grant UKRI3174.

\noindent\textit{Data availability}---The data supporting this study's findings are available within the Letter.


\begin{thebibliography}{100}

\bibitem{Freund2002}
I. Freund,  Opt. Commun. \textbf{201}, 251 (2002).

\bibitem{Zhen2014}
B. Zhen, C. W. Hsu, L. Lu, \textit{et al.}, Phys. Rev. Lett. \textbf{113}, 257401 (2014).

\bibitem{Chen2019}
A. Chen, W. Liu, Y. Zhang, \textit{et al.}, Phys. Rev. B \textbf{99}, 180101 (2019).

\bibitem{Kang2023}
M. Kang, T. Liu, C. T. Chan, \textit{et al.}, Nat. Rev. Phys. \textbf{5}, 659 (2023).

\bibitem{Tittl2018}
A. Tittl, A. Leitis, M. Liu, \textit{et al.}, Science \textbf{360}, 1105 (2018).

\bibitem{Ratiu2025}
B.-P. Ratiu, J. T. Wang, K. Netherwood, \textit{et al.}, Laser Photonics Rev. \textbf{20}, e01297 (2026).

\bibitem{Minkov2019}
M. Minkov, D. Gerace, and S. Fan, Optica \textit{6}, 1039 (2019).

\bibitem{Wang2025}
J. T. Wang and N. C. Panoiu, Rev. Phys. \textbf{13}, 100117 (2025).

\bibitem{Liu2019}
W. Liu, B. Wang, Y. Zhang, \textit{et al.}, Phys. Rev. Lett. \textbf{123}, 116104 (2019).

\bibitem{Kang2025}
M. Kang, M. Xiao, and C. T. Chan, Phys. Rev. Lett. \textbf{134}, 013805 (2025).

\bibitem{Su2026}
Z. Su, Y. Wang, B. Li, \textit{et al.}, Nano Lett. \textbf{26}, 3478 (2026).

\bibitem{Wang2026}
J. T. Wang and N. C. Panoiu, “Hybrid bound states in the continuum beyond diffraction limit,” arXiv:2601.06983 (2026).

\bibitem{Sakoda2005}
K. Sakoda, Optical Properties of Photonic Crystals, 2nd ed. (Springer, 2005).

\bibitem{Doiron2024}
C. F. Doiron, I. Brener, and A. Cerjan, Phys. Rev. Lett. \textbf{133}, 213802 (2024).

\bibitem{Cerjan2021}
A. Cerjan, C. Jorg, S. Vaidya, \textit{et al.}, Sci. Adv. \textbf{7}, eabk1117 (2021).

\bibitem{Zhang2018}
Y. Zhang, A. Chen, W. Liu, \textit{et al.}, Phys. Rev. Lett. \textbf{120}, 186103 (2018).

\bibitem{COMSOL}
COMSOL Multiphysics\textsuperscript{\textregistered}, www.comsol.com.

\bibitem{RSOFT}
Synopsys' RSoft Photonic Device Tools\textsuperscript{\textregistered}, www.synopsys.com.

\bibitem{Chong2008}
Y. D. Chong, X.-G. Wen, and M. Soljačić, Phys. Rev. B \textbf{77}, 235125 (2008).

\bibitem{Long2023}
O. Y. Long, C. Guo, and S. Fan, Phys. Rev. Appl. \textbf{20}, L051001 (2023).

\bibitem{Yoda2020}
T. Yoda and M. Notomi, Phys. Rev. Lett. \textbf{125}, 053902 (2020).

\bibitem{Jin2019}
J. Jin, X. Yin, L. Ni, \textit{et al.}, Nature \textbf{574}, 501 (2019).

\end{thebibliography}
\end{document}